\documentclass[aps,prl,twocolumn,reprint,floatfix,superscriptaddress,longbibliography]{revtex4}
\usepackage[utf8]{inputenc} 
\usepackage{graphicx}
\usepackage{amsmath}
\usepackage{amssymb}
\usepackage{amsfonts}
\usepackage{mathtools}
\usepackage{float}
\usepackage{tabularx, booktabs}
\usepackage{color}
\usepackage{braket}

\newcommand{\1}{1}
\newcommand{\2}{2}

\begin{document}
\title{Mixed bubbles in Bose-Bose mixtures}
\author{P. Naidon}
\affiliation{Strangeness Nuclear Physics Laboratory, RIKEN Nishina
Centre, Wak{\=o}, 351-0198 Japan}
\author{D. S. Petrov}
\affiliation{Universit\'e Paris-Saclay, CNRS, LPTMS, 91405, Orsay, France}

\date{\today}

\begin{abstract}

Repulsive Bose-Bose mixtures are known to either mix or phase-separate into pure components. Here we predict a mixed-bubble regime in which bubbles of the mixed phase coexist with a pure phase of one of the components. This is a beyond-mean-field effect which occurs for unequal masses or unequal intraspecies coupling constants and is due to a competition between the mean-field term, quadratic in densities, and a nonquadratic beyond-mean-field correction. We find parameters of the mixed-bubble regime in all dimensions and discuss implications for current experiments.

\end{abstract}

\maketitle

Mixtures of particles in the quantum regime have been a topic of research in various fields of physics, starting from experiments on liquid helium mixtures~\cite{Walters1956,Graf1967}. With the development of ultracold atoms, it has been possible to realise mixtures of atoms at low temperature and study their properties with full control over the system parameters such as density and interactions. In particular, a number of experiments on atomic Bose-Bose mixtures with repulsive interactions have demonstrated the phenomenon of phase separation \cite{Stenger1998,Modugno2002,Papp2008,Lee2018}, consistent with theoretical predictions based on the mean-field (MF) approximation \cite{Esry1997,Pu1998,Timmermans1998,Ao1998,Trippenbach2000,Pethick2002}.

In the MF approximation, a mixture of two components $\1$ and $\2$ is mechanically stable when the energy-density paraboloid $\sum_{\sigma\sigma'}g_{\sigma\sigma'}n_{\sigma}n_{\sigma'}/2$ is elliptic, which requires that the modulus of the interspecies coupling constant $g_{\1\2}$ be smaller than the geometrical average $\sqrt{g_{\1\1}g_{\2\2}}$ of the intraspecies coupling constants (both positive). Otherwise, the system undergoes either a phase separation for $g_{\1\2}>\sqrt{g_{\1\1}g_{\2\2}}$ or collapse for $g_{\1\2}<-\sqrt{g_{\1\1}g_{\2\2}}$. In both cases the energy density becomes a hyperbolic paraboloid, albeit the instabilities proceed in different directions in the $n_{\1}n_{\2}$-plane. One of us has shown that the beyond-mean-field (BMF) correction, not quadratic in the densities, can stabilize collapsing mixtures and can lead to their self binding~\cite{Petrov2015}. Such quantum droplets have recently been observed in potassium mixtures~\cite{Cabrera2018,Cheiney2018,Semeghini2018}, potassium-rubidium mixtures~\cite{LENSRbKDroplet} and have also been theoretically predicted to occur in lower dimensions~\cite{PetrovAstrakharchik2016}. BMF studies of mixtures close to the phase-separation threshold have focused on the stability of the mixed phase at finite temperature~\cite{Boudjemaa2018,Ota2019,Hryhorchak2019}.

In this Letter we investigate the zero-temperature BMF phases of a weakly-interacting mixture close to the miscible-immiscible threshold, i.e., for small $\delta g = g_{\1\2}-\sqrt{g_{\1\1}g_{\2\2}}$. A new feature that we predict is that for unequal intraspecies interactions or unequal masses the $\1+\2$ mixed phase can form bubbles with tunable population imbalance immersed in a pure gas of one of the components. This phenomenon is due to the competition between the MF term, artificially weakened by tuning to small $\delta g$, and the BMF correction, nonquadratic in the component densities. Mixed bubbles thus offer a relatively straightforward experimental path towards detecting quantum effects in the mixture. In spite of certain similarities, mixed bubbles differ from quantum droplets in quite a few aspects and open new avenues for further studies.

The problem is defined by the Hamiltonian
\begin{eqnarray}
\hat{H}&=&\sum_{\sigma=\1,\2}\sum_{\bf p}\frac{p^2}{2m_\sigma}\hat{a}^\dagger_{\sigma,{\bf p}}\hat{a}_{\sigma,{\bf p}}\nonumber\\
&+&\sum_{\sigma,\sigma'=\1,\2}\sum_{{\bf pqk}} \frac{g_{\sigma\sigma'}}{2} \hat{a}^\dagger_{\sigma,{\bf q}-{\bf k}} \hat{a}^\dagger_{\sigma',{\bf p}+{\bf k}} \hat{a}_{\sigma,{\bf q}} \hat{a}_{\sigma',{\bf p}},\label{Ham}
\end{eqnarray}
where we set $\hbar=1$ and assume $g_{\sigma\sigma'}$ to be constants. However, in order to avoid ultraviolet divergencies in dimensions $D=2$ and 3 the sum over ${\bf k}$ in the interaction term is cut off at $|k|>\kappa$. Then, according to the standard Bogoliubov prescription, we assume the macroscopic condensate occupations $\hat{a}_{\sigma,0}=a_{\sigma,0}$, expand Eq.~(\ref{Ham}) up to quadratic terms in the operators $\hat{a}_{\sigma,{\bf p}\neq 0}$ and $\hat{a}^\dagger_{\sigma,{\bf p}\neq 0}$, diagonalize the resulting Hamiltonian by the Bogoliubov transformation, and obtain the ground-state grand potential density in the so-called Bogoliubov approximation
\begin{equation}\label{EnBog}
\Omega=\sum_{\sigma,\sigma'}\frac{g_{\sigma\sigma'}}{2}n_\sigma n_{\sigma'}+E_{\rm B}-\mu_\1 n_1-\mu_\2 n_\2,
\end{equation}
where the Bogoliubov vacuum energy (leading BMF correction) $E_{\rm B}$ can be written explicitly (see Ref.~\cite{SM}) as a function of the masses, densities, and coupling constants. 

In order to describe the mixed-bubble effect in the most transparent fashion let us start with the mass-balanced case $m_\1=m_\2=m$, where $E_{\rm B}$ can be written in the form~\cite{Petrov2015,PetrovAstrakharchik2016}
\begin{equation}\label{Bog}
E_{\rm B}=\left\{
\begin{matrix}
\vspace{2mm}
\frac{8}{15\pi^2}\sum_{\pm}c_{\pm}^5, & D=3,\\
\vspace{4mm}
\frac{1}{8\pi}\sum_{\pm}c_{\pm}^4\ln\frac{c_{\pm}^2\sqrt{e}}{\kappa^2},& D=2,\\
-\frac{2}{3\pi}\sum_{\pm}c_{\pm}^3, & D=1,
\end{matrix}
\right.
\end{equation}
and the squared Bogoliubov sound velocities equal
\begin{equation}\label{cpm}
c_{\pm}^2=\frac{g_{\1\1}n_\1+g_{\2\2}n_\2\pm\sqrt{(g_{\1\1}n_\1-g_{\2\2}n_\2)^2+4g_{\1\2}^2 n_\1 n_\2}}{2}.
\end{equation}
Equation~(\ref{EnBog}) gives the first two leading terms in powers of the weak-interaction parameter $\gamma\ll 1$, which scales in different dimensions as $\gamma_{D=3}\propto \sqrt{m^3g^3n}$, $\gamma_{D=2}\propto mg$, and $\gamma_{D=1}\propto \sqrt{mg/n}$ (here, for estimates, we take $g_{\1\1}\sim g_{\2\2}\sim g_{\1\2}\sim g$ and $n_\1 \sim n_\2 \sim n$). We mention that in the three-dimensional case the cutoff dependence has been removed in the standard manner and in Eqs.~(\ref{EnBog}-\ref{cpm}) we use the renormalized coupling constants $g_{\sigma\sigma'}=4\pi a^{(3d)}_{\sigma\sigma'}/m$ defined by the three-dimensional scattering lengths. One can also verify \cite{PetrovAstrakharchik2016,SM} that in the two-dimensional case the grand potential is $\kappa$-independent (to the chosen approximation order) since for fixed scattering lengths $a^{(2d)}_{\sigma\sigma'}$ the coupling constants run with $\kappa$ as $g_{\sigma\sigma'}=2\pi/m\ln[2e^{\gamma}/\kappa a^{(2d)}_{\sigma\sigma'}]$, where $\gamma$ is Euler's constant. 

Since $\gamma\ll 1$ the BMF term is generally much weaker than the MF one. However, close to the phase separation threshold, in the regime $\delta g/g\sim \gamma$, they become comparable in the sense that one of the eigenvalues of the matrix $g_{\sigma\sigma'}$ is $\propto -\delta g$. The corresponding eigenvector designates a direction in the $n_{\1}n_\2$-plane, along which the system is ``soft'' and sensitive to the BMF term, whereas in the perpendicular direction the system's behavior is still governed by the dominant MF term. This separation of scales makes the analysis of the phases, which consists of minimizing Eq.~(\ref{EnBog}) with respect to the densities, a two-step process. In order to see this, let us define $g=\sqrt{g_{\1\1}g_{\2\2}}$, introduce the asymmetry parameter $\alpha=\sqrt{g_{\2\2}/g_{\1\1}}$, and rotate the $n_\1n_\2$-plane according to
\begin{eqnarray}
n_+&=&\frac{\alpha^{-1/2}n_\1+\alpha^{1/2}n_\2}{\sqrt{\alpha+\alpha^{-1}}},\label{drot}\\
n_-&=&\frac{-\alpha^{1/2}n_\1+\alpha^{-1/2}n_\2}{\sqrt{\alpha+\alpha^{-1}}},\label{srot}
\end{eqnarray}
with the constraints (equivalent to $n_\1>0$ and $n_\2>0$)
\begin{equation}\label{Interval}
n_{\rm L}<n_-<n_{\rm R},
\end{equation}
where $n_{\rm L}=-n_+\alpha$ and $n_{\rm R}=n_+/\alpha$. In these new notations the grand potential density reads
\begin{eqnarray}
\Omega&=&\frac{\alpha+\alpha^{-1}}{2}gn_+^2-\mu_+ n_+\nonumber\\
 &\hspace{-10mm}+&\hspace{-5mm}\frac{\delta g [n_+^2-(\alpha-\alpha^{-1})n_+n_--n_-^2]}{\alpha+\alpha^{-1}}+E_{\rm B}-\mu_- n_-,\label{GProtated}
\end{eqnarray}
where we introduce $\mu_+$ and $\mu_-$ given by Eqs.~(\ref{drot}) and (\ref{srot}) with $n$'s formally replaced by $\mu$'s. In Eq.~(\ref{GProtated}) we have placed the leading-order terms ($\propto gn^2$) in the first line and the next-order ones ($\propto gn^2\gamma$) in the second line. In order not to exceed the accuracy of the Bogoliubov approximation we should set $\delta g=0$ in $E_{\rm B}$ (recall that $\delta g\sim \gamma g$), which amounts to replacing $c_-$ by $0$ and $c_+^2$ by 
\begin{equation}\label{cp0}
c_+^2|_{\delta g=0}=g\frac{(\alpha^3+1)n_++\alpha(\alpha-1)n_-}{\alpha\sqrt{\alpha^2+1}}
\end{equation} 
in Eq.~(\ref{Bog}). 

Minimizing the first line of Eq.~(\ref{GProtated}) with respect to $n_+$ gives
\begin{equation}\label{dMF}
n_+=\frac{\mu_+}{g(\alpha+\alpha^{-1})},
\end{equation} 
and taking the second line into account produces a correction to Eq.~(\ref{dMF}) of order $\delta n_+\sim \gamma n_{+}$, which can be neglected in the Bogoliubov approximation [one can check that it leads to a correction $\sim gn^2\gamma^2$ in Eq.~(\ref{GProtated})]. To this order, $n_+$ is independent of the phase of the system~\cite{RemFixed}. In the second step we thus arrive at the problem of minimizing $\Omega$ with respect to $n_-$ on the interval (\ref{Interval}) with $n_+$ given by Eq.~(\ref{dMF}).

\begin{figure}
\includegraphics[clip = true, width = .8\columnwidth]{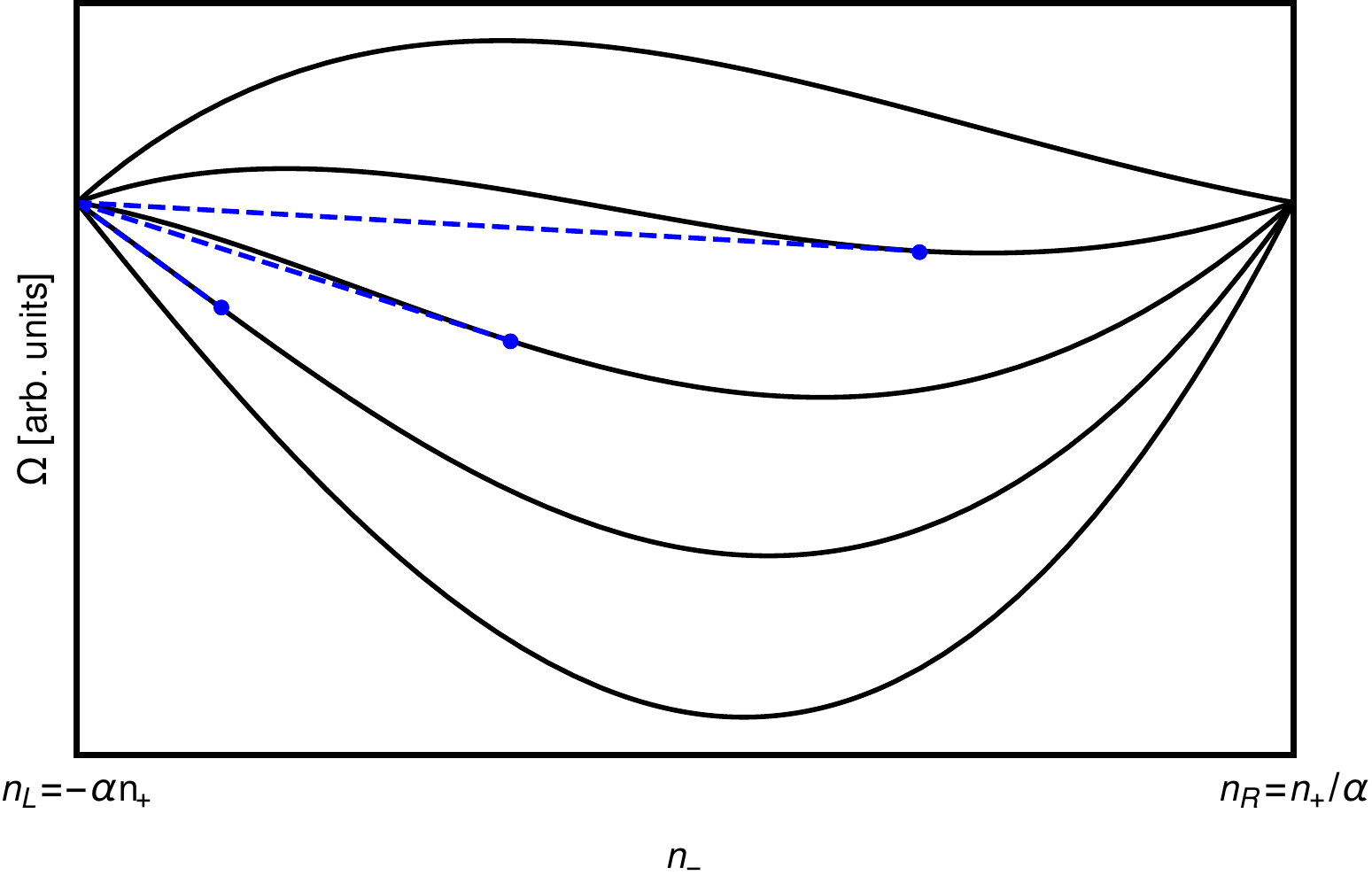}
\caption{The grand potential density $\Omega$ versus $n_-$ in the one-dimensional mass-balanced case with $\alpha=2.7$ for five values of $\delta g=\delta g_{\rm min}(1-r)+\delta g_{\rm max}r$ with (from top to bottom) $r=1.1$, 0.8, 0.5, 0.2, and -0.1. The parameters $\delta g_{\rm min}$ and $\delta g_{\rm max}$ are given by Eqs.~(\ref{gmin1D}) and (\ref{gmax1D}). For better visibility the chemical potential $\mu_-$ and a constant shift for each curve are chosen such that $\Omega$ is the same on both ends of the interval (\ref{Interval}). The dashed blue lines are the tangent constructions showing the first-order phase transitions between the pure $\1$ phase and the mixed phase.}
\label{Fig:GrandPotRen}
\end{figure}

In Fig.~\ref{Fig:GrandPotRen} we plot $\Omega(n_-)$  in the case of $D=1$ for a few values of $\delta g$, choosing $\alpha=2.7$. For sufficiently large $\delta g$ this function is concave since dominated by the term $-\delta g n_-^2/(\alpha+\alpha^{-1})$. It is thus minimized at the ends of the interval (\ref{Interval}), corresponding to the pure $\1$ and $\2$ phases. The first-order phase transition between these phases happens at $\mu_-$ defined by the equation $\Omega(n_{\rm L})=\Omega(n_{\rm R})$ (the case actually shown in Fig.~\ref{Fig:GrandPotRen}). By contrast, $\Omega(n_-)$ is convex for large negative $\delta g$ (see the lowest curve in Fig.~\ref{Fig:GrandPotRen}). In this case, the mixed phase is separated from pure phases 1 and 2 by two second-order phase transitions at $\mu_-$ determined by $\Omega'(n_{\rm L})=0$ and $\Omega'(n_{\rm R})=0$, respectively. In the MF approximation (where $E_{\rm B}=0$) these two scenarios are exhaustive; the first is realized for $\delta g>0$ and the second for $\delta g<0$. 

The BMF term $E_{\rm B}$ leads to another scenario realized for $\delta g_{\rm min}<\delta g <\delta g_{\rm max}$, where $\Omega(n_-)$ can be concave in an interval of $n_-$, and convex in another interval (see the three intermediate curves in Fig.~\ref{Fig:GrandPotRen}). For $\alpha>1$, the concave region starts at $n_{\rm L}$ (pure $\1$ phase) and ends at a certain $n_-$ inside (\ref{Interval}) corresponding to a mixed phase. The blue dashed lines in Fig.~\ref{Fig:GrandPotRen} show the tangent constructions corresponding to the first-order transitions between the pure $\1$ phase and the mixed phase (blue dots). 

In the canonical picture first-order phase transition means phase separation or, in other words, bubble formation. Component $\2$, if one tries to admix it into a big bath of $\1$ atoms, will spread over the whole system for $\delta g<\delta g_{\rm min}$. Otherwise, it will form either pure $\2$ bubbles for $\delta g>\delta g_{\rm max}$ or mixed ($\1+\2$) bubbles for $\delta g_{\rm min}<\delta g <\delta g_{\rm max}$. The consitution of this mixed bubble changes continuously from pure $\2$ to pure $\1$ phase as one decreases $\delta g$ (see the trajectory of the blue dots in Fig.~\ref{Fig:GrandPotRen}).

We note that this new scenario of mixed bubbles appears only in mixtures with unequal intraspecies interactions ($\alpha\neq 1$) since, otherwise, $c_+$ (and thus $E_{\rm B}$) does not depend on $n_-$ [see Eq.~(\ref{cp0})] and $\delta g_{\rm min}=\delta g_{\rm max}$. The effect gets enhanced with increasing $\alpha$ since $\Omega(n_-)$ then deviates more from a quadratic function. 

Although Fig.~\ref{Fig:GrandPotRen} corresponds to the concrete case $D=1$ and $\alpha=2.7$, the qualitative picture remains the same for $D>1$ and for other values of $\alpha>1$ because of the common feature that $\Omega''(n_-)$ monotonically grows with $n_-$ [one can see this by substituting Eq.~(\ref{cp0}) into Eq.~(\ref{Bog})] giving to $\Omega(n_-)$ a concave-convex look when $\delta g_{\rm min}<\delta g<\delta g_{\rm max}$. The value of $\delta g_{\rm min}$ is determined by the equation $\Omega''(n_{\rm L})=0$, which is the condition for the mixed-phase tangent point (blue dots in Fig.~\ref{Fig:GrandPotRen}) to approach the left end of the interval (\ref{Interval}). By contrast, $\delta g= \delta g_{\rm max}$ corresponds to the mixed-phase tangent point located at the right end of the interval (\ref{Interval}), which is conditioned by $\Omega'(n_{\rm R})=[\Omega(n_{\rm R})-\Omega(n_{\rm L})]/(n_{\rm R}-n_{\rm L})$. From these formulas we obtain in three dimensions
\begin{equation}\label{gmin3D}
\frac{\delta g_{\rm min}}{g}=\frac{1}{\pi^2}\frac{(\alpha-1)^2(\alpha^2+1)^{1/4}}{\alpha^{3/2}}\sqrt{m^3g^3n_{+}},
\end{equation}
\begin{equation}\label{gmax3D}
\delta g_{\rm max}=\frac{4}{15}\frac{3\alpha^{3/2}+6\alpha+4\alpha^{1/2}+2}{(\sqrt{\alpha}+1)^{2}}\delta g_{\rm min}.
\end{equation}

In two dimensions the bubble region can be defined with the help of the parameter $C$,
\begin{equation}\label{BubbleRegion2D}
2<C<\frac{1}{2}+\frac{\alpha}{\alpha-1}+\frac{(\alpha-2)\alpha\ln\alpha}{(\alpha-1)^2},
\end{equation}
related to $\delta g$ by
\begin{equation}\label{deltag2D}
\frac{\delta g}{g}=\frac{(\alpha-1)^2}{8\pi \alpha}\left[C+\ln\frac{\sqrt{\alpha^2+1}mgn_{+}}{\alpha\kappa^2}\right]mg.
\end{equation}
To give an example of application of Eq.~(\ref{deltag2D}) consider a quasi-two-dimensional mixture characterized by the three-dimensional scattering lengths $a^{(3d)}_{\sigma\sigma'}$ all much smaller than the confinement oscillator length $l$. At low energies $\ll 1/ml^2$ the two-body interaction in this geometry is equivalent to a purely two-dimensional one characterized by $g_{\sigma\sigma'}=2\sqrt{2\pi}a^{(3d)}_{\sigma\sigma'}/ml$ and $\kappa=\sqrt{\beta/\pi}/l$, where $\beta\approx 0.9$ \cite{PetrovShlyapnikov2001,Pricoupenko2008}. Equation (\ref{deltag2D}) then transforms into
\begin{equation}\label{deltaavsD}
\frac{\delta a}{a}=\frac{(\alpha-1)^2}{2\sqrt{2\pi}\alpha}\left[C+\ln\frac{(2\pi)^{3/2}\sqrt{1+\alpha^2}aln_{+}}{\alpha \beta}\right]\frac{a}{l},
\end{equation} 
where $a=\sqrt{a^{(3d)}_{\1\1}a^{(3d)}_{\2\2}}$ and $\delta a=a^{(3d)}_{\1\2}-a$.

Finally, in one dimension we have
\begin{equation}\label{gmin1D}
\frac{\delta g_{\rm min}}{g}=-\frac{1}{4\pi}\frac{(\alpha-1)^2}{\sqrt{\alpha}(\alpha^2+1)^{1/4}}\sqrt{\frac{mg}{n_{+}}},
\end{equation}
\begin{equation}\label{gmax1D}
\delta g_{\rm max}=\frac{4(\sqrt{\alpha}+2)}{3(\sqrt{\alpha}+1)^2}\delta g_{\rm min}.
\end{equation}

Note that mixed bubbles require $\delta g$ to be negative in low dimensions and positive for $D=3$. From the MF viewpoint these are, respectively, miscible and immiscible regimes. The interval $(\delta g_{\rm min},\delta g_{\rm max})$ widens with $\alpha$ and with $\gamma$. 

\begin{figure*}
\includegraphics[viewport=0bp 21bp 257bp 171bp,clip,height=39.0mm]{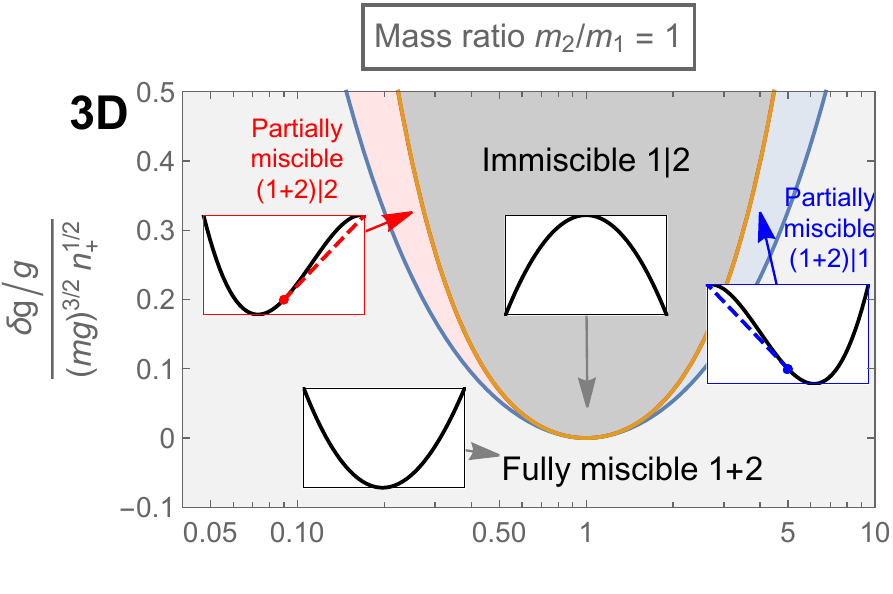}
\includegraphics[viewport=12bp 21bp 223bp 173bp,clip,height=39.0mm]{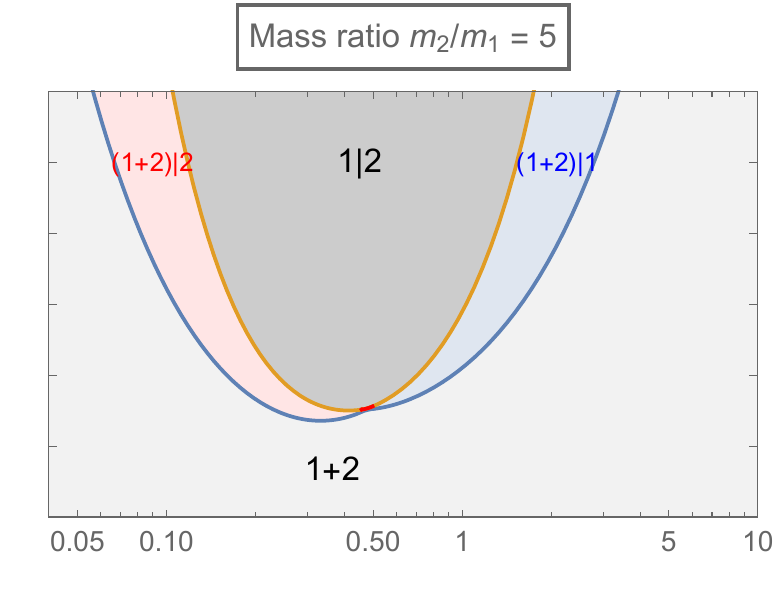}
\includegraphics[viewport=12bp 21bp 223bp 173bp,clip,height=39.0mm]{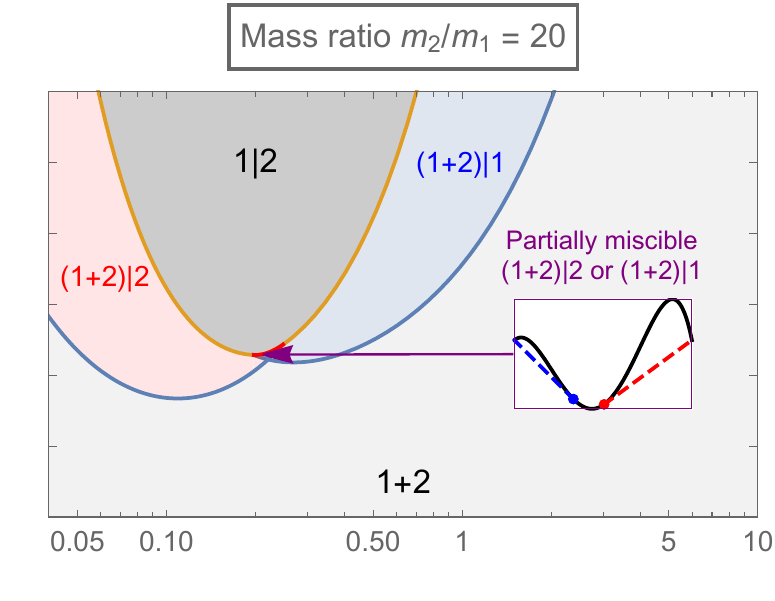}

\includegraphics[viewport=0bp 21bp 252bp 147bp,clip,height=33.63mm]{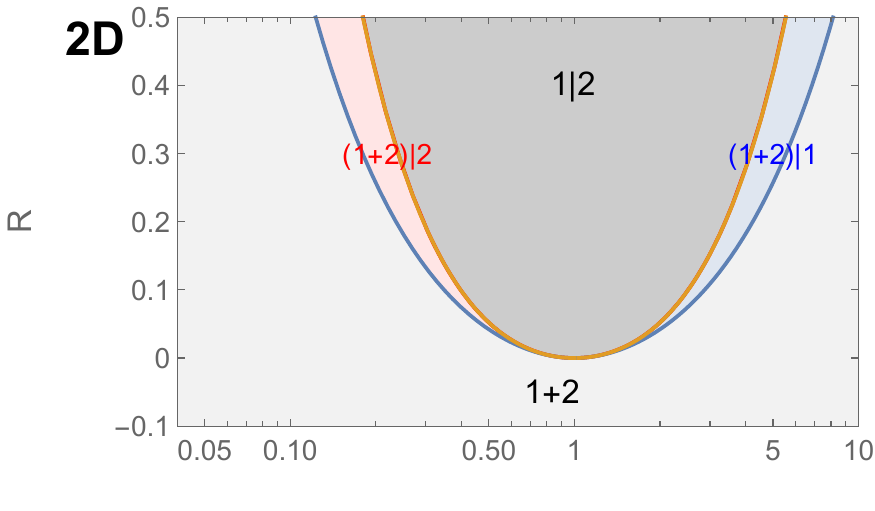}
\includegraphics[viewport=12bp 21bp 223bp 152bp,clip,height=33.63mm]{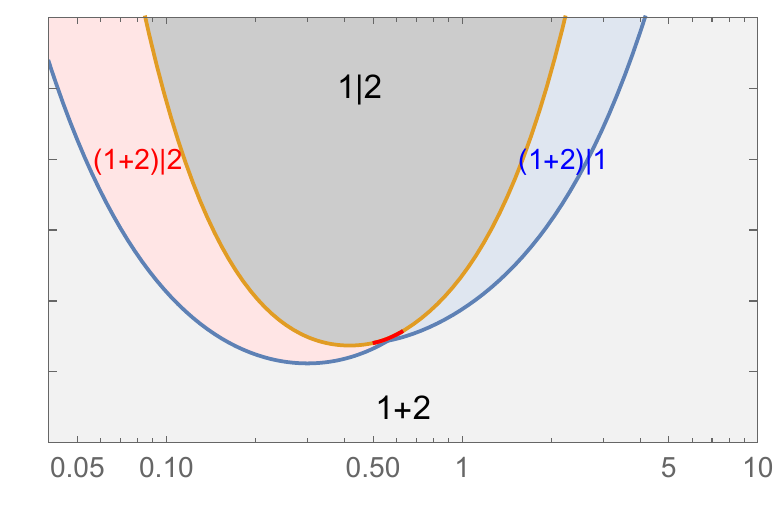}
\includegraphics[viewport=12bp 21bp 223bp 152bp,clip,height=33.63mm]{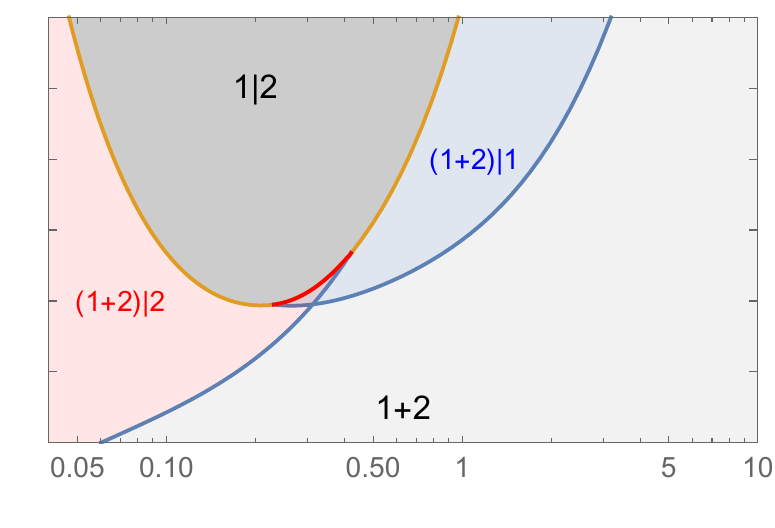}

\includegraphics[viewport=0bp 0bp 257bp 155bp,clip,height=40.4mm]{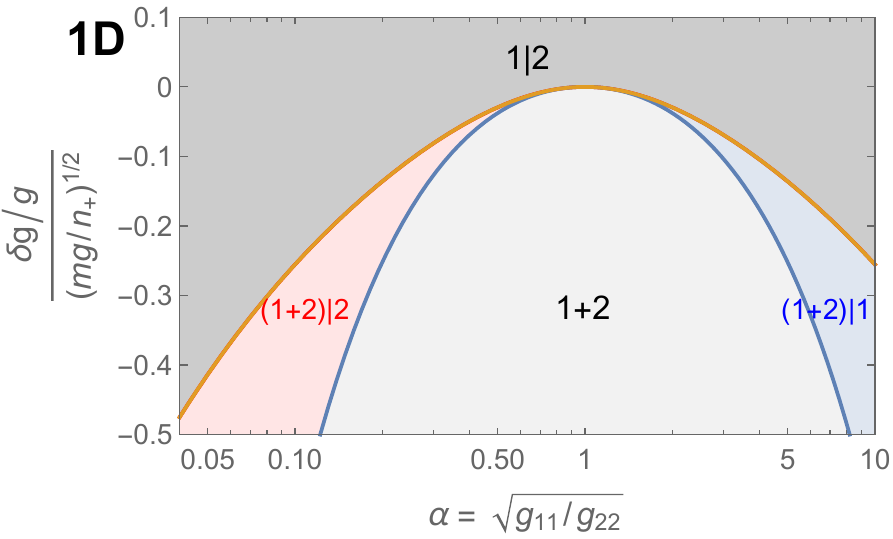}
\includegraphics[viewport=12bp 0bp 223bp 157bp,clip,height=40.4mm]{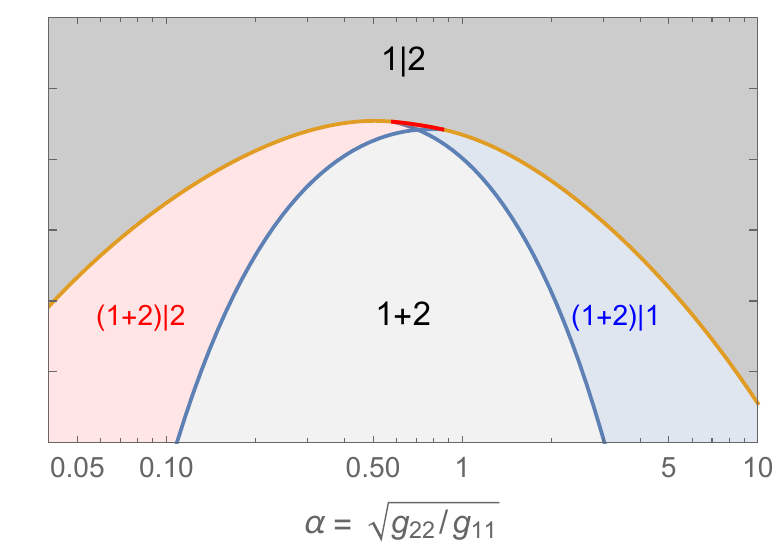}
\includegraphics[viewport=12bp 0bp 223bp 157bp,clip,height=40.4mm]{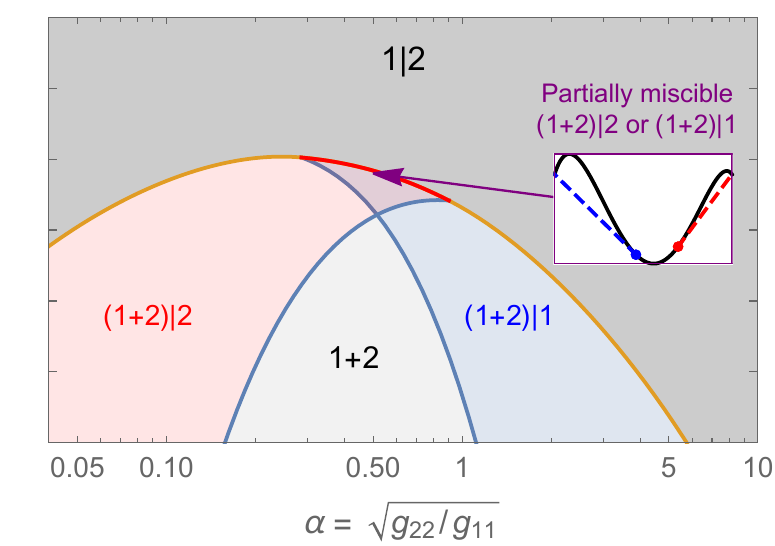}
\caption{The regions of existence of mixed bubbles in three-dimensional (upper row), two-dimensional (middle row), and one-dimensional (lower row) mixtures for three values of the mass ratio in the Bogoliubov approximation. The regions are plotted in terms of $\alpha$ and $(\delta g/g)/\gamma$ for $D=1$ and $3$. In the case $D=2$ we use $R$ defined in Eq.~(\ref{RenormalizedRatio}). The light gray areas show the miscible case and all other regions correspond to various bubble regimes. The coexistence of phases A and B in these regimes is denoted by A$|$B, where A and B stand for 1, (1+2), or 2. The regions (1+2)$|$1 and (1+2)$|$2 intersect such that mixed (1+2) bubbles can coexist there with either of the pure phases. The insets give a few examples showing the convexity of the grand potential $\Omega(n_-)$ and the tangential constructions for the parameters indicated by the arrows. 
}
\label{Fig:BubbleBoundaries}
\end{figure*}

The analysis of the mixed-bubble regime for $m_\1\neq m_\2$ in the Bogoliubov approximation is the same as in the mass-balanced case; Eqs.~(\ref{drot}), (\ref{srot}), (\ref{Interval}), (\ref{GProtated}), and (\ref{dMF}) remain valid. Although the expression for $E_{\rm B}$ is more cumbersome, it can be put in a form convenient for minimization of $\Omega(n_-)$, particularly for the relevant case $g_{\1\2}=\sqrt{g_{\1\1}g_{\2\2}}$ (see \cite{SM}). Introducing $m=\sqrt{m_\1m_\2}$ and $z=m_{\2}/m_\1$, the blue and pink areas in Fig.~\ref{Fig:BubbleBoundaries} show the mixed-bubble region in the plane $(\delta g/g)/\gamma$ versus $\alpha$ for the set of mass ratios $z=1$, 5, and 20 (from left to right) and for different dimensions $D=3,2,1$ (from top to bottom). For $D=2$, instead of $(\delta g/g)/\gamma$ we introduce
\begin{equation}\label{RenormalizedRatio}
R=\frac{\delta g}{mg^2}-\frac{1}{8\pi}\left(\frac{1}{\alpha\sqrt{z}}+\alpha\sqrt{z}-\frac{4}{\sqrt{z}+1/\sqrt{z}}\right)\ln\frac{mgn_{+}}{\kappa^2},
\end{equation}
which does not run with $\kappa$ in the Bogoliubov approximation (see more details in Ref.~\cite{SM}). Independent of $D$, in the blue-shaded regions the behavior of $\Omega(n_-)$ is qualitatively similar to the scenario depicted in Fig.~\ref{Fig:GrandPotRen}. The pink shading denotes the inverted scenario, where $\Omega(n_-)$ exhibits a convex-concave configuration and where the mixed phase can coexist with the pure $\2$ phase. For equal masses this corresponds to the exchange $\alpha\rightleftarrows 1/\alpha$ equivalent to $\1\rightleftarrows \2$. More generally, the bubble-regime boundaries for the inverse mass ratios ($z=1$, 1/5, and 1/20) can be obtained from Fig.~\ref{Fig:BubbleBoundaries} by replacing $\alpha\rightarrow 1/\alpha$ and exchanging the blue and pink shading.

In Fig.~\ref{Fig:BubbleBoundaries} we see that the mixed-bubble region significantly widens with increasing the mass imbalance. This feature, promising from the viewpoint of observing the mixed bubbles, is due to the amplification of the nonquadratic part of $E_{\rm B}$, particularly when $\ln\alpha$ and $\ln z$ are of the same sign. A noticeable peculiarity of the mass-imbalanced cases is that when $\ln\alpha$ and $\ln z$ are of different signs, effects of the mass and interaction imbalance compete with each other and pinch the mixed-bubble region. However, since the two effects cannot completely cancel the nonquadratic part of $E_{\rm B}$, the pinch areas (curved triangles in Fig.~\ref{Fig:BubbleBoundaries}) are, in fact, realizations of yet another scenario where $\Omega(n_-)$ acquires a concave-convex-concave configuration and allows for two separate tangential constructions (see the insets in right upper and lower panels in Fig.~\ref{Fig:BubbleBoundaries}). Note that this scenario becomes more probable with decreasing $D$.

Quite a few currently available ultracold mixtures are suitable for the observation of mixed bubbles. For instance, the mixture of $^{41}$K (component $\1$) and $^{39}$K (component $\2$) atoms both in hyperfine states $F=1$, $m_F=0$, is characterized by a $\2-\2$ Feshbach resonance at $B\approx 60$G~\cite{DErrico2007}, in the vicinity of which the other scattering lengths equal $a^{(3d)}_{\1\1}\approx 65a_0$ and $a^{(3d)}_{\1\2}\approx 174a_0$~\cite{Tanzi2018}. Neglecting the mass imbalance, the MF miscible-immiscible threshold is thus achieved by tuning $a^{(3d)}_{\2\2}$ to the value $(a^{(3d)}_{\1\2})^2/a^{(3d)}_{\1\1}\approx 470 a_0$ corresponding to $\alpha\approx 2.7$ (explaining our choice of $\alpha$ in Fig.~\ref{Fig:GrandPotRen}). Another example, the $^{174}$Yb-$^{7}$Li mixture studied in Ref.~\cite{Schaefer2018}, is among the most mass imbalanced. This mixture can be tuned near the MF miscible-immiscible threshold at $B\approx 650$G thanks to the $^{7}$Li resonance at $B\approx 700$G.

In contrast to self-trapped liquid droplets, mixed bubbles are pockets trapped inside a gaseous medium, which requires an external trapping. However, the trap should be sufficiently flat in order not to interfere with the subtle MF-BMF competition at the heart of the mixed-bubble physics. We leave this point for future studies.
Other open questions are the shape of finite-size bubbles, their dynamics, excitation spectra, and superfluid properties. Again in contrast to the droplet case, bubble characteristics should depend on the velocity with which they move through the host gas. This may become a route towards probing Andreev-Bashkin physics \cite{FilShevchenko2005} and other BMF effects. That the mixed-bubble region widens with $\gamma$ suggests further studies of strongly-interacting regimes, particularly for $D=1$, where large $\gamma$ is not generally associated with enhanced losses.

We thank L. Tarruell, Y. Takahashi, and F. Sch\"afer for useful discussions and communications. PN acknowledges support from the RIKEN Incentive Research Project and JSPS Grants-in-Aid for Scientific Research on Innovative Areas (No. JP18H05407). DSP acknowledges support from ANR Grant Droplets No. ANR-19-CE30-0003-02 and hospitality of IQOQI Innsbruck where part of this research has been conducted.

\newpage

\renewcommand{\theequation}{S\arabic{equation}}
\renewcommand{\thefigure}{S\arabic{figure}}

\setcounter{equation}{0}
\setcounter{figure}{0}

\begin{widetext}

\centerline{{\bf Supplemental Material: Mixed bubbles in Bose-Bose mixtures}}
\vspace{5mm}
\centerline{P.~Naidon and D.~S.~Petrov}

\section{The Bogoliubov vacuum energy $E_{\rm B}$ (Lee-Huang-Yang term)}

\subsection{Case $D=3$}

In three dimensions we have 
\begin{equation}\label{EB3dSM}
E^{(3d)}_{\rm B}=\frac{8}{15\pi^2}\left(\frac{m_\1}{\hbar^2}\right)^{3/2}(g_{\1\1}n_\1)^{5/2}f^{(3d)}\left(\frac{m_\2}{m_\1},\frac{g_{\1\2}^2}{g_{\1\1}g_{\2\2}},\frac{g_{\2\2}n_\2}{g_{\1\1}n_\1}\right),
\end{equation}
where
\begin{eqnarray}
f^{(3d)}(z,u,x) &=&\frac{15}{32}
\int_0^\infty \left[\frac{1}{\sqrt{2}}\sum_{\pm}\sqrt{k^2+\frac{xk^2}{z}+\frac{k^4}{4} + \frac{k^4}{4z^2} 
\pm \sqrt{\left(k^2-\frac{xk^2}{z}+\frac{k^4}{4}-\frac{k^4}{4z^2}\right)^2 +  \frac{4xuk^4}{z}}}\right.\nonumber\\
&&\hspace{6cm}\left.-\;\frac{1+z}{2z}\, k^2 -1-x+\left(1+x^2z+\frac{4xzu}{1+z}\right)\frac{1}{k^2}\right] k^2dk.\label{f3dSM}
\end{eqnarray}

In order to calculate the integral we change the integration variable by using
\begin{equation}\label{ktotSM}
k^2=\frac{4\sqrt{xuz^3}}{z^2-1}\left[t-\frac{1}{t}+\frac{x-z}{\sqrt{xuz}}\right]=\frac{4\sqrt{xuz^3}}{z^2-1}\frac{(t-b_1)(t-b_2)}{t}.
\end{equation}
Assuming $z>1$ and $b_2>b_1$, the new integration interval is $t\in [b_2,\infty)$. The change of variable (\ref{ktotSM}) removes the internal square root in Eq.~(\ref{f3dSM}) leading to
\begin{equation}\label{identitypSM}
\sqrt{k^2+\frac{xk^2}{z}+\frac{k^4}{4} + \frac{k^4}{4z^2} 
+ \sqrt{\left(k^2-\frac{xk^2}{z}+\frac{k^4}{4}-\frac{k^4}{4z^2}\right)^2 +  \frac{4xuk^4}{z}}}=\frac{\sqrt{8xuz}}{z^2-1}\frac{\sqrt{(t-b_1)(t-b_2)(t-a_1)(t-a_2)}}{t}
\end{equation}
and
\begin{equation}\label{identitymSM}
\sqrt{k^2+\frac{xk^2}{z}+\frac{k^4}{4} + \frac{k^4}{4z^2} 
- \sqrt{\left(k^2-\frac{xk^2}{z}+\frac{k^4}{4}-\frac{k^4}{4z^2}\right)^2 +  \frac{4xuk^4}{z}}}=\frac{z\sqrt{8xuz}}{z^2-1}\frac{\sqrt{(t-b_1)(t-b_2)(t-a_1/z^2)(t-a_2/z^2)}}{t},
\end{equation}
where $a_1$ and $a_2$ are roots of 
\begin{equation}\label{a1a2SM}
t^2+\frac{\sqrt{z}(xz-1)}{\sqrt{xu}}t-z^2=0.
\end{equation}
The integration of Eq.~(\ref{f3dSM}) then results in a combination of elementary and elliptic functions. 
In the particular case $g_{\1\2}=\pm\sqrt{g_{\1\1}g_{\2\2}}$ we obtain 
\begin{eqnarray}\label{fMassImb3dSM}
f^{(3d)}(z,1,x)&=&(-2 - 7 x z + 2 z^2 + x^2 z^2)\frac{\sqrt{x+z}}{2\sqrt{z}(z^2-1)}\nonumber\\
& &+ (-2 - 7 x z + 3 z^2 + 3 x^2 z^2 - 7 x z^3 - 2 x^2 z^4)\frac{E[{\rm arcsin} (1/z)|-xz]-E(-xz)}{2(z^2-1)^{3/2}}\nonumber\\
& & + (2 + 8 x z - 3 z^2 + 6 x^2 z^2 - 2 x z^3 + x^2 z^4)\frac{F[{\rm arcsin} (1/z)|-xz]-K(-xz)]}{2(z^2-1)^{3/2}}.
\end{eqnarray}
In Eq.~(\ref{fMassImb3dSM}) $E(\phi|v)$ is the elliptic integral of the second kind, $E(v)$ is the complete elliptic integral, $F(\phi|v)$ is the elliptic integral of the first kind, and $K(v)$ is the complete elliptic integral of the first kind. 

Finally, let us mention the identity $f^{(3d)}(z,u,x)=z^{3/2}x^{5/2}f^{(3d)}(1/z,u,1/x)$ which follows from Eq.~(\ref{EB3dSM}) and which can be useful, for instance, for analyzing the vicinity of the extreme limit $n_\1\rightarrow 0$, where $x$ diverges.  

\subsection{Case $D=2$}

In two dimensions we have 
\begin{equation}\label{EB2dSM}
E^{(2d)}_{\rm B}=\frac{1}{4\pi}\left(\frac{m_\1}{\hbar^2}\right)(g_{\1\1}n_\1)^{2}f^{(2d)}\left(\frac{m_\2}{m_\1},\frac{g_{\1\2}^2}{g_{\1\1}g_{\2\2}},\frac{g_{\2\2}n_\2}{g_{\1\1}n_\1},\frac{\kappa}{\sqrt{m_\1 g_{\1\1} n_\1}}\right),
\end{equation}  
where
\begin{equation}\label{f2dSM}
f^{(2d)}(z,u,x,\tilde{\kappa}) =
\int_0^{\tilde{\kappa}} \left[\frac{1}{\sqrt{2}}\sum_{\pm}\sqrt{k^2+\frac{xk^2}{z}+\frac{k^4}{4} + \frac{k^4}{4z^2} 
\pm \sqrt{\left(k^2-\frac{xk^2}{z}+\frac{k^4}{4}-\frac{k^4}{4z^2}\right)^2 +  \frac{4xuk^4}{z}}}-\frac{1+z}{2z}\, k^2 -1-x\right] kdk.
\end{equation}
This function satisfies $f^{(2d)}(z,u,x,\tilde{\kappa})=zx^2f^{(2d)}(1/z,u,1/x,\tilde{\kappa}/\sqrt{zx})$ and, at large $\tilde{\kappa}$, can be written as
\begin{equation}\label{Asympf2SM}
f^{(2d)}(z,u,x,\tilde{\kappa}) =-\frac{1 + z + 4 u x z + x^2 z + x^2 z^2}{1 + z}\ln\tilde{\kappa}+\eta(z,u,x)+O(\tilde{\kappa}^{-2}).
\end{equation}
We neglect the effective-range correction $O(\tilde{\kappa}^{-2})$, which is exponentially small in terms of the expansion parameter $\gamma$. For our analysis it is sufficient to set $g_{\1\2}=\pm\sqrt{g_{\1\1}g_{\2\2}}$ and we make use of the explicit expression
\begin{eqnarray}\label{fMassImb2dSM}
\eta(z,1,x)&=&\frac{x^2z^2+x^2z+4xz+z+1}{z+1}\ln\frac{2x^{1/4}z^{3/4}}{\sqrt{z^2-1}}+\frac{x^2z^3+4x^2z^2-x^2z+z^2-4z-1}{4(z^2-1)}-\frac{(xz-1)\sqrt{z(x+z)(1+xz)}}{z^2-1}\nonumber\\
& -&\frac{x^2z^5-2x^2z^3+4xz^4-4xz^3+z^4+x^2z-4xz^2+4xz-2z^2+1}{(z^2-1)^2}\ln 2 -\frac{4xz^3-z^4-4xz+2z^2-1}{(z^2-1)^2}\ln\frac{z-1}{x^{1/4}z^{3/4}} \nonumber\\
&  +&\frac{xz(xz^2+4z-x)}{z^2-1}\ln\frac{(x/z)^{3/4}(z-1)^{3/2}\sqrt{z+1}}{\sqrt{1+xz}-\sqrt{1+x/z}}-\frac{4xz-z^2+1}{2(z^2-1)}\ln\frac{z\sqrt{1+x/z}+\sqrt{1+xz}}{z\sqrt{1+x/z}-\sqrt{1+xz}}.
\end{eqnarray}
and the property (useful for the case $z<1$)
\begin{equation}\label{symmetry2dSM}
\eta(z,1,x)=x^2z\eta(1/z,1,1/x)+\frac{x^2z^2+x^2z+4xz+z+1}{z+1}\ln\sqrt{xz}.
\end{equation}

\subsection{Case $D=1$}

In one dimension we have 
\begin{equation}\label{EB1dSM}
E^{(1d)}_{\rm B}=\frac{1}{2\pi}\left(\frac{m_\1}{\hbar^2}\right)^{1/2}(g_{\1\1}n_\1)^{3/2}f^{(1d)}\left(\frac{m_\2}{m_\1},\frac{g_{\1\2}^2}{g_{\1\1}g_{\2\2}},\frac{g_{\2\2}n_\2}{g_{\1\1}n_\1}\right),
\end{equation}  
where
\begin{equation}\label{f1dSM}
f^{(1d)}(z,u,x) =
\int_0^\infty \left[\frac{1}{\sqrt{2}}\sum_{\pm}\sqrt{k^2+\frac{xk^2}{z}+\frac{k^4}{4} + \frac{k^4}{4z^2} 
\pm \sqrt{\left(k^2-\frac{xk^2}{z}+\frac{k^4}{4}-\frac{k^4}{4z^2}\right)^2 +  \frac{4xuk^4}{z}}}-\frac{1+z}{2z}\, k^2 -1-x\right] dk
\end{equation}
satisfies $f^{(1d)}(z,u,x)=z^{1/2}x^{3/2}f^{(1d)}(1/z,u,1/x)$. In the particular case $g_{\1\2}=\pm\sqrt{g_{\1\1}g_{\2\2}}$ we use the explicit expression
\begin{eqnarray}\label{fMassImb1dSM}
f^{(1d)}(z,1,x)&=&-\frac{4}{3}\sqrt{1+\frac{x}{z}}+\frac{4}{3}\frac{1}{\sqrt{z^2-1}}\biggl\{(xz-1)E[{\rm arcsin} (1/z)|-xz]+(xz+1)F[{\rm arcsin} (1/z)|-xz]\biggr.\nonumber\\
&&\hspace{-2cm}\biggl.-2(xz-1)[E(-xz)-(xz+1)K(-xz)]+\sqrt{xz+1}\left[(xz-1)E\left(\frac{xz}{xz+1}\right)-(2xz-1)K\left(\frac{xz}{xz+1}\right)\right]\biggr\}.
\end{eqnarray}

\section{Mixed-bubble regime boundaries}

\subsection{Case $D=3$}

In the upper row of Fig.~\ref{Fig:BubbleBoundaries} of the main text, the blue and orange curves comprising the blue areas are determined, respectively, from the conditions $\Omega''(n_{\rm L})=0$ and $\Omega'(n_{\rm R})=[\Omega(n_{\rm R})-\Omega(n_{\rm L})]/(n_{\rm R}-n_{\rm L})$, the same conditions as in the mass-balanced case, explained in the main text. They give, respectively, 
\begin{equation}\label{deltagmin3dSM}
\frac{\delta g_{\rm min}(z,\alpha)}{g\sqrt{m^3g^3n_+}}=\frac{(\alpha^2+1)^{1/4}}{\pi^2\alpha^{3/2}z^{3/4}}\left[\frac{-4 + 4 z^2 + 12 z \alpha + 3 z^2 \alpha^2}{4(z^2-1)}+\frac{3z^2\alpha(-4z-2\alpha+z^2\alpha)}{4(z^2-1)^{3/2}}{\rm arccos}(1/z)\right]
\end{equation}
and 
\begin{equation}\label{deltagmax3dSM}
\frac{\delta g_{\rm max}(z,\alpha)}{g\sqrt{m^3g^3n_+}}=\frac{(\alpha^2+1)^{1/4}}{\pi^2\alpha^{3/2}z^{3/4}}\left[\frac{-8 + 8 z^2 - 30 z^{5/2} \alpha^{3/2} - 12 z^{3/2} \alpha^{5/2} + 
 12 z^{7/2} \alpha^{5/2}}{15(z^2-1)}-\frac{2(z\alpha)^{3/2}\ln (z-\sqrt{z^2-1})}{(z^2-1)^{3/2}}\right].
\end{equation}

The boundaries comprising the pink areas are given, respectively, by Eqs.~(\ref{deltagmin3dSM}) and (\ref{deltagmax3dSM}) with replaced $\alpha\rightarrow 1/\alpha$ and $z\rightarrow 1/z$ (equivalent to the interchange $\1\rightleftarrows \2$).

The red upper borders of the ``pinch'' areas are obtained by the condition that the convex lobe of $\Omega(n_-)$ has a minimum degenerate with $\Omega(n_{\rm L})$ and $\Omega(n_{\rm R})$.

\subsection{Case $D=2$}

The two-dimensional case is analyzed in the same manner as the three-dimensional one. In spite of the seemingly explicit dependence on $\kappa$ the results are actually cut-off independent (to the chosen approximation order) because of the running
\begin{equation}\label{gvskappaSM}
g_{\sigma\sigma'}=\frac{\pi}{\mu_{\sigma\sigma'}\ln[2e^\gamma/\kappa a_{\sigma\sigma'}^{(2d)}]},
\end{equation}
where $\mu_{\sigma\sigma'}=m_\sigma m_{\sigma'}/(m_\sigma+m_{\sigma'})$. In our notations ($z=m_\2/m_\1$ and $m=\sqrt{m_\1 m_\2}$) $\mu_{\1\1}=m/2\sqrt{z}$, $\mu_{\2\2}=m\sqrt{z}/2$, and $\mu_{\1\2}=m/(\sqrt{z}+1/\sqrt{z})$. Equation~(\ref{gvskappaSM}) is valid for $\gamma=\mu_{\sigma\sigma'}g_{\sigma\sigma'}\ll 1$, which gives the freedom to choose the cut-off scale. Indeed, choosing another cutoff $\tilde{\kappa}$ the new coupling constant $\tilde{g}_{\sigma\sigma'}$ is related to the old one by the equation
\begin{equation}\label{grunning}
g_{\sigma\sigma'}\approx \tilde{g}_{\sigma\sigma'}-\frac{\mu_{\sigma\sigma'}\tilde{g}^2_{\sigma\sigma'}}{\pi}\ln\frac{\tilde{\kappa}}{\kappa}\approx \tilde{g}_{\sigma\sigma'}-\frac{\mu_{\sigma\sigma'}g^2_{\sigma\sigma'}}{\pi}\ln\frac{\tilde{\kappa}}{\kappa}.
\end{equation}  
The tilde is removed in the second-order term since the difference between $g^2$ and $\tilde{g}^2$ is third order in $\gamma$ and can be neglected in the Bogoliubov approximation. In the same spirit the $\kappa$-dependence of $g_{\sigma\sigma'}$ in the MF energy term gets cancelled by the explicit logarithmic dependence of $E_{\rm B}$.  

From Eq.~(\ref{grunning}) one can also derive the running of $\delta g=g_{\1\2}-\sqrt{g_{\1\1}g_{\2\2}}$. Namely,
\begin{equation}\label{deltagrunning}
\delta{g}\approx \delta{\tilde{g}}+\frac{m\tilde{g}^2}{4\pi}\left(\frac{1}{\alpha\sqrt{z}}+\alpha\sqrt{z}-\frac{4}{\sqrt{z}+1/\sqrt{z}}\right)\ln\frac{\tilde{\kappa}}{\kappa}\approx \delta{\tilde{g}}+\frac{mg^2}{4\pi}\left(\frac{1}{\alpha\sqrt{z}}+\alpha\sqrt{z}-\frac{4}{\sqrt{z}+1/\sqrt{z}}\right)\ln\frac{\tilde{\kappa}}{\kappa}.
\end{equation} 
The middle row in Fig.~\ref{Fig:BubbleBoundaries} thus presents the boundaries of the mixed-bubble regime in terms of the renormalized ratio
\begin{equation}\label{RenormalizedRatioSM}
R=\frac{\delta g}{mg^2}-\frac{1}{8\pi}\left(\frac{1}{\alpha\sqrt{z}}+\alpha\sqrt{z}-\frac{4}{\sqrt{z}+1/\sqrt{z}}\right)\ln\frac{mgn_{+}}{\kappa^2},
\end{equation}
which is $\kappa$ independent in the Bogoliubov approximation. The blue and orange curves comprising the blue-shaded regions in the middle row of Fig.~\ref{Fig:BubbleBoundaries} are obtained in the same manner as Eqs.~(\ref{deltagmin3dSM}) and (\ref{deltagmax3dSM}) and are given, respectively, by the formulas
\begin{eqnarray}\label{RminSM}
R_{\rm min}(z,\alpha)&=&\frac{1}{16\pi(z^2-1)\sqrt{z}\alpha}[(2+4\ln 2)z^3\alpha^2+4z^2\alpha^2-(8+16\ln 2)z^2\alpha-(6+4\ln 2)z\alpha^2+8z\alpha+4z^2-4\nonumber\\
&+&(3z^3\alpha^2-12z^2\alpha-3z\alpha^2-z^2-4z\alpha+1)\ln z - (2z^3\alpha^2-8z^2\alpha-2z\alpha^2+2z^2+8z\alpha-2)\ln\alpha\nonumber\\
&+&(z^3\alpha^2-4z^2\alpha-z\alpha^2+z^2+4z\alpha-1)\ln(\alpha^2+1)-(4z^3\alpha^2-16z^2\alpha-4z\alpha^2)\ln(z+1). 
\end{eqnarray}
and
\begin{equation}\label{RmaxSM}
R_{\rm max}(z,\alpha)=\frac{3z\alpha^2+1-\ln[z\alpha^2/(\alpha^2+1)]}{16\pi\sqrt{z}\alpha}+\frac{\sqrt{z}(z\alpha+\alpha-4)\ln [z(\alpha^2+1)]}{16\pi(z+1)}-\frac{\sqrt{z}\ln [(z+1)/2]}{\pi(z^2-1)}. 
\end{equation}
The pink-shaded areas are restricted by the curves $R_{\rm min}(1/z,1/\alpha)$ (blue) and  $R_{\rm max}(1/z,1/\alpha)$ (orange). The red boundary is calculated numerically from the condition that $\Omega(n_-)$ has three degenerate minima.

\subsection{Case $D=1$}

The analysis is also the same. The blue and orange curves for the blue-shaded regions in the lower row of Fig.~\ref{Fig:BubbleBoundaries} are obtained from the formulas,
\begin{equation}\label{deltagmin1dSM}
\frac{\delta g_{\rm min}(z,\alpha)}{g\sqrt{mg/n_+}}=-\frac{1}{8\pi z^{1/4} \sqrt{\alpha}(\alpha^2+1)^{1/4}}\left[2+\alpha^2+\frac{z\alpha(z\alpha-4)}{\sqrt{z^2-1}}{\rm arccos}(1/z)\right]
\end{equation}
and 
\begin{equation}\label{deltagmax1dSM}
\frac{\delta g_{\rm max}(z,\alpha)}{g\sqrt{mg/n_+}}=-\frac{1}{3\pi[z\alpha^2(\alpha^2+1)]^{1/4}}\left[2+\sqrt{z\alpha^3}+\frac{3\sqrt{z\alpha}\ln(z-\sqrt{z^2-1})}{\sqrt{z^2-1}}\right],
\end{equation}
and, as in higher dimensions, the pink-shaded region boundaries are obtained from Eqs.~(\ref{deltagmin1dSM}) and (\ref{deltagmax1dSM}) by setting $\alpha\rightarrow 1/\alpha$ and $z\rightarrow 1/z$.

\end{widetext}


\begin{thebibliography}{99}

\bibitem{Walters1956}
G.~K. Walters and W.~M. Fairbank, Phase Separation in ${\mathrm{He}}^{3}$---${\mathrm{He}}^{4}$ Solutions, Phys. Rev. {\bf 103},  262 (1956).

\bibitem{Graf1967}

E.~H. Graf, D.~M. Lee, and J.~D. Reppy, Phase Separation and the Superfluid
  Transition in Liquid ${\mathrm{He}}^{3}$-${\mathrm{He}}^{4}$ Mixtures, Phys. Rev. Lett. {\bf 19},  417 (1967).


\bibitem{Stenger1998}

J.~Stenger, S.~Inouye, D.~M. Stamper-Kurn, H.-J. Miesner, A.~P. Chikkatur, and
  W.~Ketterle, Spin domains in ground-state Bose-Einstein condensates, Nature {\bf 396},  345 (1998).


\bibitem{Modugno2002}

G.~Modugno, M.~Modugno, F.~Riboli, G.~Roati, and M.~Inguscio, Two Atomic
  Species Superfluid, Phys. Rev. Lett. {\bf 89},  190404 (2002).


\bibitem{Papp2008}

S.~B. Papp, J.~M. Pino, and C.~E. Wieman, Tunable Miscibility in a
  Dual-Species Bose-Einstein Condensate, Phys. Rev. Lett. {\bf 101},  040402 (2008).


\bibitem{Lee2018}

K.~L. Lee, N.~B. J{\o}rgensen, L.~J. Wacker, M.~G. Skou, K.~T. Skalmstang,
  J.~J. Arlt, and N.~P. Proukakis, Time-of-flight expansion of binary
  Bose{\textendash}Einstein condensates at finite temperature, New Journal of
  Physics {\bf 20},  053004 (2018).


\bibitem{Esry1997}

B.~D. Esry, C.~H. Greene, J.~P. Burke, Jr., and J.~L. Bohn, Hartree-Fock
  Theory for Double Condensates, Phys. Rev. Lett. {\bf 78},  3594 (1997).


\bibitem{Pu1998}

H.~Pu and N.~P. Bigelow, Properties of Two-Species Bose Condensates, Phys. Rev. Lett. {\bf 80},  1130 (1998).


\bibitem{Timmermans1998}

E.~Timmermans, Phase Separation of Bose-Einstein Condensates, Phys. Rev. Lett. {\bf 81},  5718 (1998).


\bibitem{Ao1998}

P.~Ao and S.~T. Chui, Binary Bose-Einstein condensate mixtures in weakly and
  strongly segregated phases, Phys. Rev. A {\bf 58},  4836 (1998).


\bibitem{Trippenbach2000}

M.~Trippenbach, K.~G{\'{o}}ral, K.~Rzazewski, B.~Malomed, and Y.~B. Band,
  Structure of binary Bose-Einstein condensates, J. Phys. B {\bf 33},  4017 (2000).


\bibitem{Pethick2002}

C.~J.~Pethick and H.~Smith, \emph{Bose-Einstein Condensation in Dilute Gases}, Cambridge University Press, (2002).

\bibitem{Boudjemaa2018}
A.~Boudjem\^{a}a, Quantum and thermal fluctuations in two-component Bose gases, Phys. Rev. A {\bf 97} 033627 (2018).

\bibitem{Ota2019}
M.~Ota, S.~Giorgini, and S.~Stringari, Phys. Rev. Lett. {\bf 123}, 075301 (2019).

\bibitem{Hryhorchak2019}
O. Hryhorchak, V. Pastukhov, Large-N expansion for condensation and stability of Bose-Bose mixtures at finite temperatures, arXiv:1904.12351 (2019).

\bibitem{Petrov2015}
D.~S.~Petrov, Quantum mechanical stabilization of a collapsing Bose-Bose mixture, Phys. Rev. Lett. {\bf 115}, 155302 (2015).

\bibitem{Cabrera2018}
C.~R.~Cabrera, L.~Tanzi, J.~Sanz, B.~Naylor, P.~Thomas, P.~Cheiney, and L.~Tarruell, Quantum liquid droplets in a mixture of Bose-Einstein condensates, Science {\bf 359}, 301 (2018).

\bibitem{Cheiney2018}
P.~Cheiney, C.~R.~Cabrera, J.~Sanz, B.~Naylor, L.~Tanzi, and L.~Tarruell, Bright Soliton to Quantum Droplet Transition in a Mixture of Bose-Einstein Condensates, Phys. Rev. Lett. {\bf 120}, 135301 (2018).

\bibitem{Semeghini2018}
G.~Semeghini, G.~Ferioli, L.~Masi, C.~Mazzinghi, L.~Wolswijk, F.~Minardi, M.~Modugno, G.~Modugno, M.~Inguscio, and M.~Fattori, Self-Bound Quantum Droplets of Atomic Mixtures in Free Space, Phys. Rev. Lett. {\bf 120}, 235301 (2018).

\bibitem{LENSRbKDroplet}
C.~D'Errico, A.~Burchianti, M.~Prevedelli, L.~Salasnich, F.~Ancilotto, M.~Modugno, F.~Minardi, and C.~Fort, Observation of quantum droplets in a heteronuclear bosonic mixture, Phys. Rev. Research {\bf 1}, 033155 (2019).

\bibitem{PetrovAstrakharchik2016}
D.S. Petrov and G.E. Astrakharchik, Ultradilute low-dimensional liquids, Phys. Rev. Lett. {\bf 117}, 100401 (2016).

\bibitem{RemFixed}
The analog of such a fixed quantity in the case of quantum droplets is the population ratio $n_\1/n_\2=\sqrt{g_\2/g_\1}$, which stays approximately constant while the total density changes.  

\bibitem{DErrico2007}
C.~D'Errico, M.~Zaccanti, M.~Fattori, G.~Roati, M.~Inguscio, G.~Modugno, and A.~Simoni, Feshbach resonances in ultracold $^{39}$K, New J. Phys. {\bf 9}, 223 (2007).

\bibitem{Tanzi2018}
L.~Tanzi, C.~R. Cabrera, J.~Sanz, P.~Cheiney, M.~Tomza, and L.~Tarruell, Feshbach resonances in potassium Bose-Bose mixtures, Phys. Rev. A {\bf 98}, 062712 (2018).

\bibitem{Schaefer2018}
F.~Sch\"afer, N.~Mizukami, P.~Yu, S.~Koibuchi, A.~Bouscal, and Y.~Takahashi, Experimental realization of ultracold Yb-$^{7}\mathrm{Li}$ mixtures in mixed dimensions, Phys. Rev. A, {\bf 98}, 051602 (2018).

\bibitem{SM}
Supplemental Material contains explicit expressions for $E_{\rm B}$, $\delta g_{\rm min}$, and $\delta g_{\rm max}$ in the mass-imbalanced cases.


\bibitem{PetrovShlyapnikov2001}
D.~S.~Petrov and G.~V.~Shlyapnikov, Interatomic collisions in a tightly confined Bose gas, Phys. Rev. A {\bf 64}, 012706 (2001).

\bibitem{Pricoupenko2008}
L.~Pricoupenko, Resonant Scattering of Ultracold Atoms in Low Dimensions, Phys. Rev. Lett. {\bf 100}, 170404 (2008).

\bibitem{FilShevchenko2005}
D.~V.~Fil and S.~I.~Shevchenko, Nondissipative drag of superflow in a two-component Bose gas, Phys. Rev. A {\bf 72}, 013616 (2005).


\end{thebibliography}
\end{document}